\documentclass[11pt,preprint]{aastex}

\usepackage{amsmath}        
\usepackage{amssymb}

\shorttitle{Radio emission in Aql X-1}
\shortauthors{Miller-Jones et al.}

\begin{document}

\title{Evolution of the radio - X-ray coupling throughout an entire outburst of Aquila X-1}

\author{J. C. A. Miller-Jones\altaffilmark{1,15}, G. R. Sivakoff\altaffilmark{2}, D. Altamirano\altaffilmark{3}, V. Tudose\altaffilmark{4,16,17}, S. Migliari\altaffilmark{5}, V. Dhawan\altaffilmark{6}, R. P. Fender\altaffilmark{7,3}, M. A. Garrett\altaffilmark{4,18,19}, S. Heinz\altaffilmark{8}, E. G. K\"ording\altaffilmark{9}, H. A. Krimm\altaffilmark{10,20}, M. Linares\altaffilmark{11}, D. Maitra\altaffilmark{12}, S. Markoff\altaffilmark{3}, Z. Paragi\altaffilmark{13,21}, R. A. Remillard\altaffilmark{11}, M. P. Rupen\altaffilmark{6}, A. Rushton\altaffilmark{7}, D. M. Russell\altaffilmark{3}, C. L. Sarazin\altaffilmark{2}, R. E. Spencer\altaffilmark{14}}
\email{jmiller@nrao.edu}

\altaffiltext{1}{NRAO Headquarters, 520 Edgemont Road, Charlottesville, VA 22902.}
\altaffiltext{2}{Department of Astronomy, University of Virginia, P.O. Box 400325, Charlottesville, VA 22904}
\altaffiltext{3}{Astronomical Institute `Anton Pannekoek', University of Amsterdam, P.O. Box 94249, 1090 GE Amsterdam, the Netherlands}
\altaffiltext{4}{Netherlands Institute for Radio Astronomy, Oude Hoogeveensedijk 4, 7991 PD Dwingeloo, the Netherlands}
\altaffiltext{5}{European Space Astronomy Centre, Apartado/P.O. Box 78,
  Villanueva de la Canada, E-28691 Madrid, Spain}
\altaffiltext{6}{NRAO Domenici Science Operations Center, 1003 Lopezville Road, Socorro, NM 87801}
\altaffiltext{7}{School of Physics and Astronomy, University of Southampton, Southampton SO17 1BJ, UK}
\altaffiltext{8}{Astronomy Department, University of Wisconsin-Madison, 475. N. Charter St., Madison, WI 53706}
\altaffiltext{9}{Universit\'e Paris Diderot and Service d'Astrophysique, UMR AIM, CEA Saclay, F-91191 Gif-sur-Yvette, France}
\altaffiltext{10}{NASA/Goddard Space Flight Center, Greenbelt, MD 20771}
\altaffiltext{11}{MIT Kavli Institute for Astrophysics and Space Research, Building 37, 70 Vassar Street, Cambridge, MA 02139}
\altaffiltext{12}{Department of Astronomy, University of Michigan, Ann Arbor, MI 48109}
\altaffiltext{13}{Joint Institute for VLBI in Europe, Postbus 2, 7990 AA Dwingeloo, the Netherlands}
\altaffiltext{14}{Jodrell Bank Centre for Astrophysics, School of Physics and
  Astronomy, University of Manchester, Manchester M13 9PL, UK}
\altaffiltext{15}{Jansky Fellow}
\altaffiltext{16}{Astronomical Institute of the Romanian Academy, Cutitul de Argint 5, RO-040557 Bucharest, Romania}
\altaffiltext{17}{Research Center for Atomic Physics and Astrophysics, Atomistilor 405, RO-077125 Bucharest, Romania}
\altaffiltext{18}{Leiden Observatory, University of Leiden, PO Box 9513, 2300 RA Leiden, the Netherlands}
\altaffiltext{19}{Centre for Astrophysics and Supercomputing, Swinburne University of Technology, Hawthorn, 3122 Victoria, Australia}
\altaffiltext{20}{Universities Space Research Association, Columbia, MD}
\altaffiltext{21}{MTA Research Group for Physical Geodesy and Geodynamics, PO Box 91, H-1521 Budapest, Hungary}

\begin{abstract}
The 2009 November outburst of the neutron star X-ray binary Aquila X-1
was observed with unprecedented radio coverage and simultaneous
pointed X-ray observations, tracing the radio emission around the full
X-ray hysteresis loop of the outburst for the first time.  We use
these data to discuss the disc-jet coupling, finding the radio
emission to be consistent with being triggered at state transitions,
both from the hard to the soft spectral state and vice versa.  Our
data appear to confirm previous suggestions of radio quenching in the
soft state above a threshold X-ray luminosity of $\sim10$\% of the
Eddington luminosity.  We also present the first detections of Aql X-1
with Very Long Baseline Interferometry (VLBI), showing that any
extended emission is relatively diffuse, and consistent with steady
jets rather than arising from discrete, compact knots.  In all cases
where multi-frequency data were available, the source radio spectrum
is consistent with being flat or slightly inverted, suggesting that
the internal shock mechanism that is believed to produce optically
thin transient radio ejecta in black hole X-ray binaries is not active
in Aql X-1.

\end{abstract}

\keywords{ X-rays: binaries --- radio continuum: stars --- stars: individual
  (Aql X-1) --- astrometry }

\section{Introduction}

Multi-wavelength observations of accreting stellar-mass compact
objects have revealed a fundamental coupling between the processes of
accretion and ejection.  As an X-ray binary (XRB) evolves through a
set of characteristic X-ray spectral and variability states over the
course of an outburst, the radio emission, assumed to arise from jets
launched from the inner regions of the system, is also seen to
evolve. \citet{Fen04} proposed a phenomenological model for this
accretion-ejection (disc-jet) coupling in the relatively well-studied
black hole (BH) XRBs.  In this model, compact steady jets are present
in the hard X-ray spectral state, with the jet power scaling with
X-ray luminosity.  The jet velocity increases when the source makes a
transition to a softer spectral state, leading to internal shocks
within the flow \citep[e.g.][]{Kai00}, observed as discrete knots of
radio emission that move outwards at relativistic speeds.  The compact
jets then switch off.  The system eventually makes a transition back to
a hard spectral state, whereupon the core jets are re-established, and
the source fades back into quiescence.  This second state transition
tends to occur at a lower luminosity than the original transition from
the hard to the soft state, implying a hysteresis in the outburst
cycle \citep{Mac03}.

Compact jets are also inferred to exist in neutron star (NS) systems,
from brightness temperature arguments, the observed flat-spectrum
radio emission, and the detection of a jet break in broadband spectra
\citep{Mig06b,Mig10}.  Rapid energy transfer from the core to detached
radio lobes \citep{Fom01,Fen04} also argues strongly for the existence
of jets in NS systems, although their compact jets have not been
directly imaged, as in BH systems \citep{Dha00,Sti01}.  Since NS XRBs
are typically fainter radio emitters than their BH counterparts at the
same X-ray luminosity \citep{Fen01,Mig06}, the nature of the disc-jet
coupling in NS systems is consequently less well understood.
\citet{Mig06} made a systematic study of radio emission from NS XRBs,
finding evidence for steady jets in hard state systems at low
luminosities ($<1$\% of the Eddington luminosity, $L_{\rm Edd}$) and
transient jets in outbursting sources close to $L_{\rm Edd}$, as seen
in BH systems but without complete suppression of the radio emission
in soft states.  The correlation between X-ray and radio emission in
hard states was steeper in NS systems \citep{Mig06} than the analogous
correlation in BH systems \citep{Gal03}, implying a different coupling
between X-ray luminosity and mass accretion rate, although the jet
power for a given accretion rate is similar in both classes of system
\citep{Koe06}.

Understanding the similarities and differences between the disc-jet
coupling in BH and NS systems is crucial in determining the role
played by the depth of the gravitational potential well, the stellar
surface, and any stellar magnetic field in the process of jet
formation.

\subsection{Aql X-1}

Aql X-1 is a recurrent transient XRB which undergoes outbursts every
$\sim300$\,d.  The accretor is a confirmed NS \citep{Koy81}, in a
$\sim19$-h orbit \citep{Che91,Sha98,Wel00} with a K7V companion
\citep{Che99}.  Its X-ray spectral and timing behavior classify it as
an atoll source \citep{Rei00}.  The distance, determined from the
luminosity of its Type I X-ray bursts, is in the range 4.4--5.9\,kpc
\citep{Jon04}.

Despite its frequent outbursts, there are few reported radio
detections of Aql X-1, likely owing to the faintness of atoll sources
in the radio band \citep{Mig06}.  \citet{Tud09} analyzed all
publicly-available archival data from the Very Large Array (VLA),
taken exclusively during X-ray outbursts, between 1986 and 2005. They
detected the source at 11 epochs, with a maximum radio flux density of
0.4\,mJy.  In all cases where multi-frequency observations were
available, the radio spectrum was inverted ($\alpha>0$, where flux
density $S_{\nu}$ scales with frequency $\nu$ as
$S_{\nu}\propto\nu^{\alpha}$).  They also found tentative evidence for
quenching of the radio emission above a certain X-ray luminosity.

Here, we extend on the work of \citet{Tud09}, presenting the most
complete radio coverage of an outburst of Aql X-1 obtained to date, to
elucidate the nature of the radio/X-ray coupling in this source.

\section{Observations}

Rising X-ray flux from Aql X-1 was observed on 2009 November 1
\citep{Lin09} by the Proportional Counter Array (PCA) on board the
RXTE satellite during the ongoing Galactic bulge monitoring program
\citep{Mar00}.  Here we present follow-up radio and X-ray data taken
during the outburst.

\subsection{VLA}

On detection of rising X-ray flux, we triggered VLA monitoring at
8.4\,GHz.  Data were taken in dual circular polarization with a total
observing bandwidth of 100\,MHz.  The array was in its least-extended
D configuration.  On detection of radio emission \citep{Siv09}, we
also began observing at 4.8\,GHz until the source faded below our
detection threshold.  Further, we reduced all remaining unpublished
archival VLA data, covering the outbursts of 2006 August and 2007
October.  Data reduction was carried out according to standard
procedures within AIPS \citep{Gre03}.

\subsection{VLBA}

Following the initial VLA radio detection, we triggered observations
with the Very Long Baseline Array (VLBA).  We observed at 8.4\,GHz in
dual circular polarization, using the maximum available recording rate
of 512\,Mbps, corresponding to an observing bandwidth of 64\,MHz per
polarization.  The observations, of duration 2--6\,h, were
phase-referenced to the nearby calibrator J1907+0127, from the third
extension to the VLBA Calibrator Survey \citep[VCS-3;][]{Pet05} and
located $1.33^{\circ}$ from Aql X-1.  The phase-referencing cycle time
was 3\,min, with occasional scans on the VCS-4 \citep{Pet06} check
source J1920-0236.  For the longer observations, 30\,min at the start
and end of the observing run were used to observe bright calibrator
sources at differing elevations to calibrate unmodeled clock and
tropospheric phase errors, thereby improving the success of the phase
transfer.  Data reduction was carried out according to standard
procedures within AIPS.

\subsection{EVN}

As reported by \citet{Tud09b}, Aql X-1 was observed at 5 GHz with the
European VLBI Network (EVN) on 2009 November 19 (14:30--19:00 UT)
using the e-VLBI technique. The array comprised stations at
Effelsberg, Medicina, Onsala, Torun, Westerbork, Yebes and Cambridge
with a maximum recording rate of 1 Gbps.  We observed in
full-polarization mode with a total bandwidth of 128 MHz per
polarization, and data were phase referenced to J1907+0127. The data
were calibrated in AIPS and imaged in Difmap \citep{She97} using
standard procedures.

\subsection{RXTE}
\label{sec:rxte}

Following \citet{Alt08}, we used the 16-s time-resolution Standard 2
mode PCA data to calculate X-ray colors and intensities.  Hard and
soft colors were defined as the count rate ratios
(9.7--16.0~keV~/~6.0--9.7~keV) and (3.5--6.0~keV~/~2.0--3.5~keV),
respectively, and intensity was defined as the 2.0--16.0~keV count
rate.  Type I X-ray bursts were removed, background was subtracted and
deadtime corrections were made, before normalizing intensities and
colors by those of the Crab Nebula, on a per-PCU basis.

\section{Results}

\subsection{Lightcurves}
\label{sec:lc}

\begin{figure}[t]
\epsscale{0.8}
\plotone{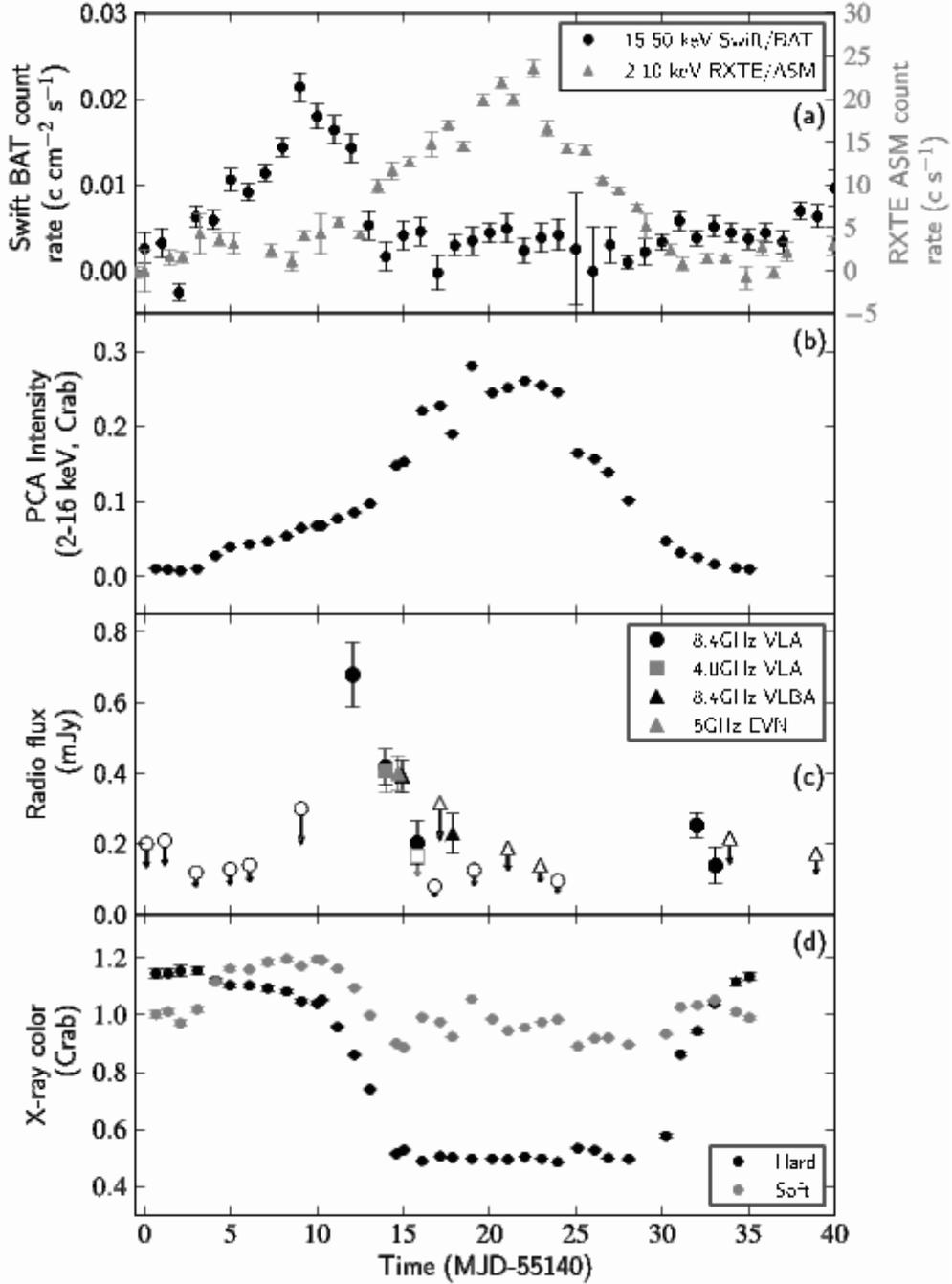}
\caption{Radio and X-ray lightcurves of the 2009 November outburst of
  Aql X-1.  (a) Swift BAT count rate\protect\footnotemark[1] in black
  (left-hand axis) and RXTE ASM count rate\protect\footnotemark[2] in
  grey (right-hand axis).  (b) RXTE PCA intensity.  (c) Radio flux
  density.  Circles and squares indicate VLA observations and
  triangles VLBI observations.  Black and grey markers denote 8.4 and
  5\,GHz observations, respectively.  Filled markers denote detections
  and corresponding open markers the $3\sigma$ upper limits.  (d)
  X-ray colors, defined as in Section \ref{sec:rxte}.  Note the radio
  detections at the state transitions where the hard X-ray color
  changes.
  \label{fig:lcs}}
\end{figure}

The observational results are given in Table \ref{tab:obs} and plotted
in Fig.~\ref{fig:lcs}, together with the publicly available 2--10\,keV
RXTE ASM and 15--50\,keV Swift BAT lightcurves.  The hard X-ray flux
peaks first during the outburst, as noted by \citet{Yu03}.  As this
starts to decrease, the soft flux rises and the X-ray colors decrease.
Radio emission is first detected at this point of X-ray spectral
softening, with an 8.4 GHz flux density of $0.68\pm0.09$\,mJy (the
highest level measured to date from Aql X-1).  The hard X-ray flux
then levels off while the radio emission fades below detectable levels
and the soft X-ray flux peaks at $\sim300$\,mCrab before decaying.  As
the X-ray color increases again and the spectrum hardens, a second
burst of radio emission is detected.  Thus while the radio sampling is
relatively sparse during the second state transition, the radio
emission is consistent with being triggered at transitions in both
directions between the hard and the soft X-ray states.

\subsection{Imaging}

\begin{figure}
\epsscale{0.8}
\plotone{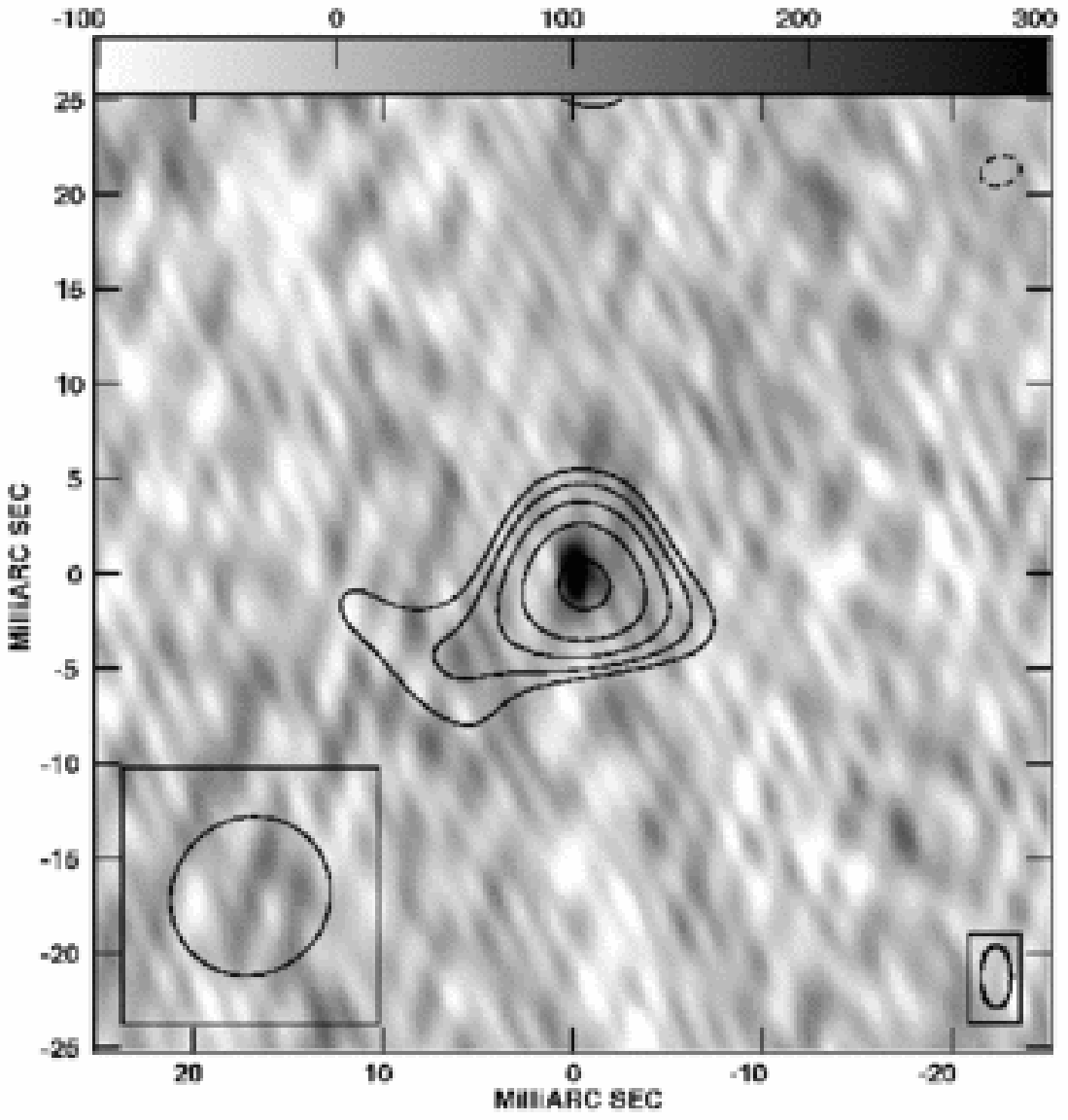}
\caption{Greyscale 8.4-GHz VLBA image of Aql X-1 from 2009 November 19
  (MJD 55154--55155), with 5-GHz EVN contours overlaid.  Contours are
  at levels of $\pm(\sqrt{2})^n$ times the rms noise level of
  37\,microJy/beam, where $n=2,3,4...$.  EVN and VLBA beamsizes are
  shown at lower left and right, respectively.  Note the marginal
  ($\sim3\sigma$) extension to the southeast in the EVN image. No
  corresponding extension is seen with the VLBA.
  \label{fig:images}}
\end{figure}

During the X-ray spectral softening, the source was detected for the
first time with VLBI.  The 5-GHz EVN observations were not fully
consistent with an unresolved source, showing tentative evidence for a
marginal-significance ($\sim3\sigma$, $0.08$\,mJy) extension to the
southeast (Fig.~\ref{fig:images}).  The 8.4-GHz VLBA observations made
on the same day were consistent with an unresolved source, to a noise
level of 0.046\,mJy\,beam$^{-1}$, and implied a slightly inverted core
radio spectrum ($\alpha=0.40\pm0.31$).  Within uncertainties, the
VLBI flux density was consistent at both frequencies with the
integrated VLA flux density measured the previous day, so the majority
of the flux was recovered on milliarcsecond scales.  However, faint
extended emission at the level seen by the EVN cannot be fully
discounted, since convolving the 8.4-GHz VLBA data to the 5-GHz EVN
resolution raised the noise level too high to detect such emission.
Simulations demonstrated that the longer baselines of the VLBA would
not detect faint extended emission as seen by the EVN on a size scale
of 10\,mas unless it were significantly more compact than the EVN
beamsize.  This allows for the possibility that the emission is
sufficiently faint and extended to be resolved out by the VLBA
(possibly indicative of a large-scale, steady jet), but makes it
unlikely to be caused by compact, localized internal shocks.
Alternatively, the extension in the EVN image may not in fact be real.

The VLBA position of Aql X-1, relative to the VCS-3 phase
reference source J\,1907+0127, whose position was taken to be
(J\,2000) 19$^{\rm h}$07$^{\rm m}$11\fs9962510(747)
01\degr27$^{\prime}$08\farcs96251(135), was
\begin{equation}
\begin{split}
\notag
{\rm RA} &= 19^{\rm h}11^{\rm m}16\fs0251654 \pm 0.000006\\
{\rm Dec.} &= +00\degr35^{\prime}05\farcs8920 \pm 0.0002\qquad{\rm (J\,2000)},
\end{split}
\end{equation}
where the quoted error bars are purely statistical errors from fitting
in the image plane.  Imaging the check source J\,1920-0236, phase
referenced to J\,1907+0127, suggested that the systematic errors are
well below 1\,mas.  This VLBA position is also consistent within
errors with the EVN position.

\section{Discussion and comparison with black hole systems}

The hardness-intensity diagram (HID) is a diagnostic plot used
extensively in interpreting the X-ray evolution of BH XRB outbursts
\citep[e.g.][]{Fen04}.  \citet{Mai04} found that the 2000 September
outburst of Aql X-1 traced out a very similar hysteresis loop in a HID
to the better-studied BH candidates.  Fig.~\ref{fig:hid} shows the
radio emission superposed on the HIDs for both the 2009 November
outburst of Aql X-1 and previous outbursts, showing the relation
between radio emission and the X-ray spectral state of the source.

\subsection{The relation between X-ray state and radio emission}

\begin{figure*}
\epsscale{1.0}
\plottwo{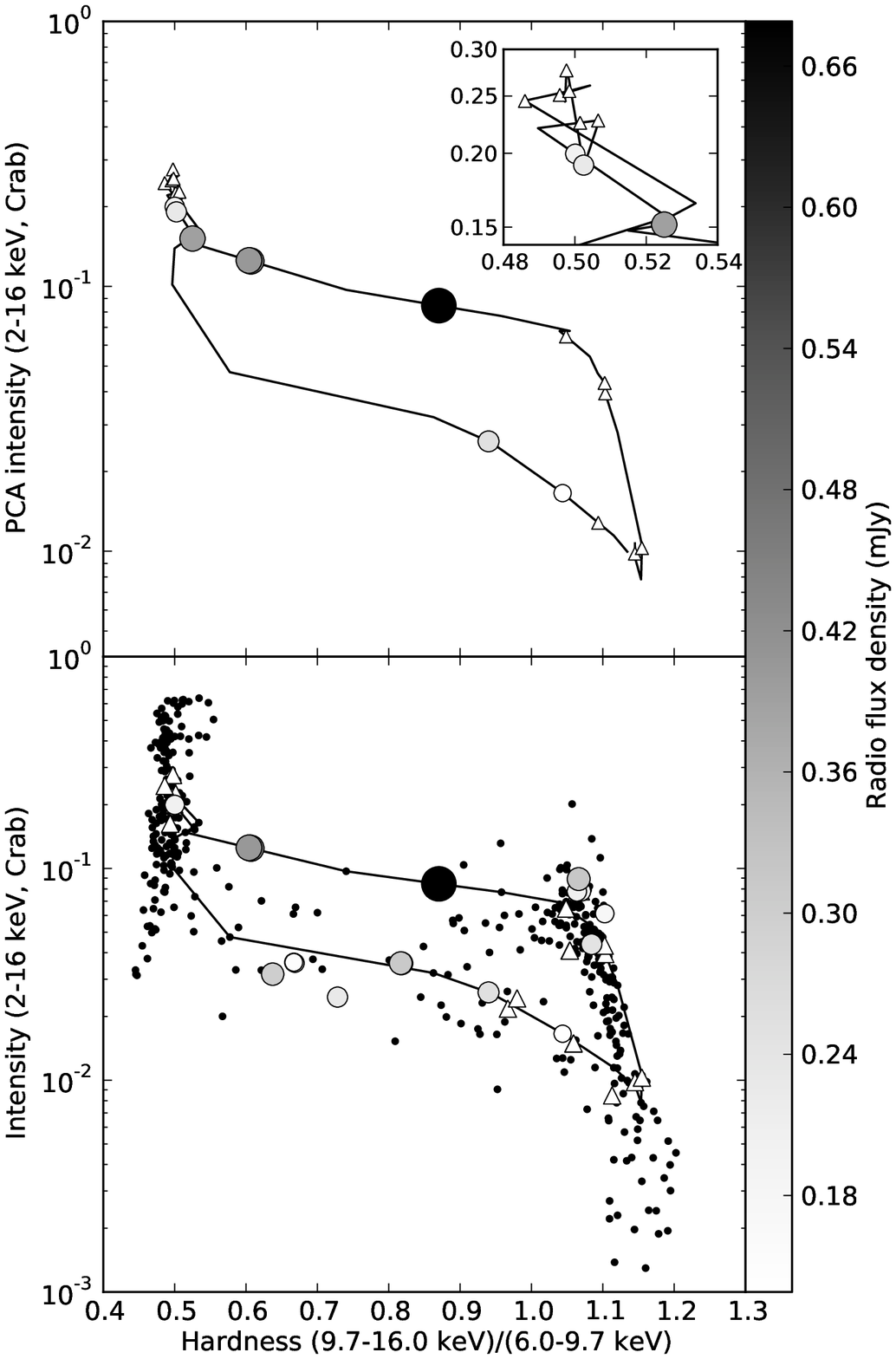}{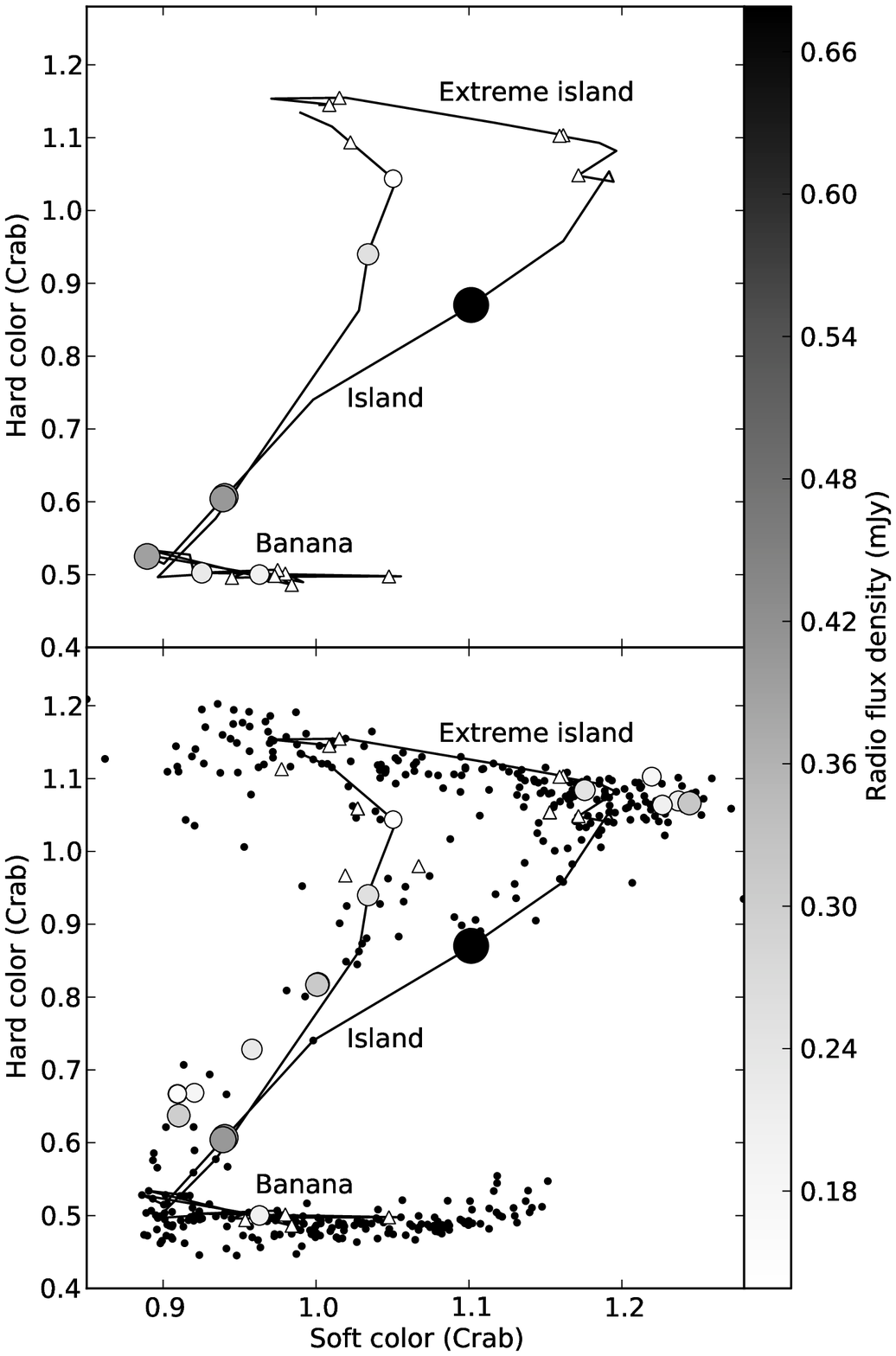}
\caption{Left: HIDs for the 2009 November outburst of Aql X-1 (top)
  and all recorded outbursts with archival PCA data (bottom).  Right:
  Color-color diagrams (CCDs) for the 2009 November outburst (top) and
  the archival outbursts (bottom). During the 2009 outburst, the
  source moves anticlockwise around the HID hysteresis loop and
  clockwise around the CCD track (solid lines in both figures).  The
  X-ray state at the time of the radio observations has been
  interpolated from X-ray observations within 2\,d.  Circles indicate
  radio detections at either 5 or 8.4\,GHz, with the VLA, VLBA or EVN.
  The size and color of the circles represent the radio flux density,
  according to the scale shown on the right of the plots.  Open
  triangles denote radio upper limits.  Dots show RXTE PCA
  measurements without simultaneous radio observations.  The inset in
  the top HID shows a zoomed-in version of the soft state (upper left
  corner of the HID).  Radio data from previous outbursts are taken
  from \citet{Tud09} and Table \ref{tab:obs}.  These plots show how
  the radio emission corresponds to the X-ray state of the source.
  \label{fig:hid}}
\end{figure*}

The radio emission is consistent with being triggered at X-ray state
transitions, switching on as the source moves into the intermediate
(island) state in the color-color diagram (CCD; Fig.~\ref{fig:hid}),
from either the hard (extreme island) or soft (banana) state.  This
pattern of triggering radio emission at state transitions is similar
to the behavior seen in both the atoll source 4U\,1728-34
\citep{Mig03} and in BH XRB outbursts \citep{Fen04}.  However, unlike
in BH systems, we did not detect radio emission in the rising hard
state (the right hand side of the HID), possibly due to insufficient
sensitivity \citep[our upper limits are close to the predicted values
  from the correlation of][]{Mig06}, nor did we detect optically-thin
shocked ejecta after the transition to the soft state.

\begin{figure}
\epsscale{1.}
\plotone{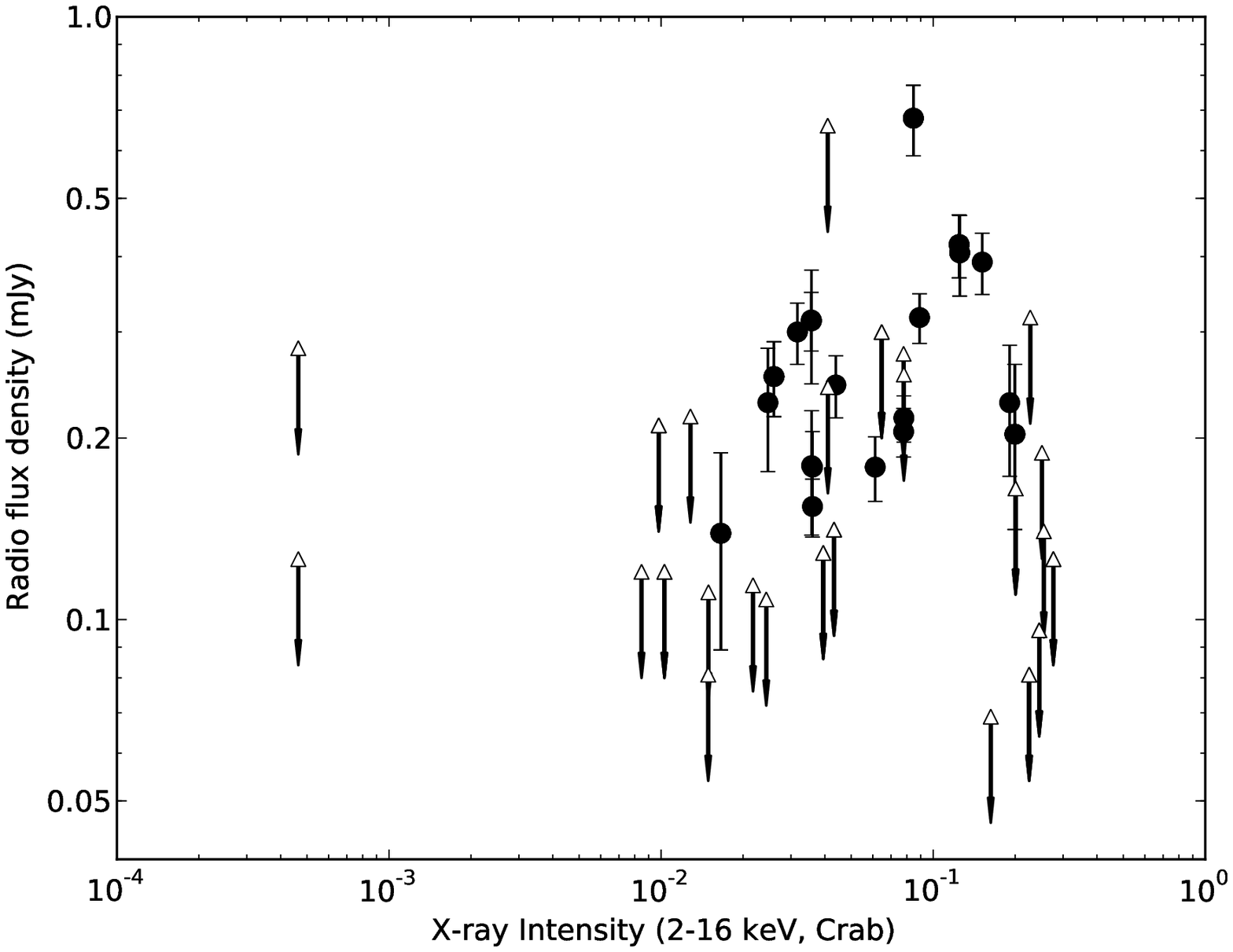}
\caption{Correlation between radio and X-ray emission for all observations of Aql X-1, regardless of X-ray state.  Note the quenching of the radio emission at high X-ray luminosities.
  \label{fig:rxcorr}}
\end{figure}

The radio emission appears quenched while the system is in a
high-luminosity soft (banana) state (Fig.~\ref{fig:rxcorr}).
\citet{Mig03} and \citet{Tud09} also found evidence for the quenching
of the jets in NS systems above a critical X-ray luminosity, albeit on
the basis of relatively sparse observational data \citep[but see][for
  counterexamples]{Mig04}.  From Figures~\ref{fig:hid} and
\ref{fig:rxcorr}, the critical luminosity for Aql X-1 appears to be
$\sim0.21$ Crab, corresponding to a 2--16\,keV luminosity of
$1.8\times10^{37}(d/5{\rm kpc})^2$\,erg\,s$^{-1}$, i.e.\ $0.1L_{\rm
  Edd}$ for a $1.4M_{\odot}$ neutron star accretor.  While reminiscent
of the drop in radio flux seen in BH systems at a few per cent of
$L_{\rm Edd}$ \citep{Gal03}, higher-sensitivity observations are
required to accurately determine the critical luminosity.

\subsection{The nature of the jets}

The brightness temperature ($T_{\rm B}$) of the unresolved VLBA source
is $T_{\rm B}>4.3\times10^6$\,K. The maximum VLA flux density of
0.68\,mJy implies a source with $r=1.0\times10^4 T_{\rm B}^{-1/2}
(d/5{\rm kpc})$\, AU. Even at the inverse Compton limit of $T_{\rm
  B}=10^{12}$\,K, the emission would arise on scales greater than the
Roche lobe around the NS, which strongly suggests that the emitting
material is flowing away from the system.

The two epochs with constraints on the radio spectrum are consistent
with flat or slightly-inverted spectra, typically interpreted in BH
XRBs as arising from partially self-absorbed compact jets
\citep{Bla79}.  Reduction of archival VLA data \citep[Table
  \ref{tab:obs} and][]{Tud09} showed that every dual-frequency
detection of the source to date is consistent with a flat or inverted
radio spectrum, albeit with fairly large error bars.  Reducing the
error bars by taking a weighted mean of all spectral indices taken
outside the hard (extreme island) X-ray state (in which, by analogy
with BH, a flat spectrum is expected) gives $\alpha=0.05\pm0.16$.
Such a flat spectrum is also consistent with the lack of
optically-thin ejecta in the VLBI observations, and with the recovery
of the full integrated VLA flux with the VLBA and EVN.  We thus find
no evidence for transient, optically thin jet emission as seen in BH
XRB outbursts.

If confirmed by future observations, the marginal EVN extension would
imply a jet of size $\sim10$\,mas ($<50(d/5{\rm kpc})$\,AU).  If not, the
VLBA beamsize of $2.9\times1.1$\,mas$^2$ gives an upper limit on the
physical size scale of the jets of $<15(d/5{\rm kpc})$\,AU at
8.4\,GHz.  Such a small size is consistent with the low flux density,
when compared to the case of Cyg X-1 \citep{Hei06}.

Only two NS XRBs, Sco X-1 and Cir X-1, have previously been imaged on
VLBI scales.  \citet{Fom01} detected resolved radio lobes in Sco X-1,
which moved out from the central source at speeds of 0.3--0.6$c$.
Rather than being internal shocks within a jet, these were interpreted
as the working surfaces where the jets impacted the surrounding
environment.  From the time delay between corresponding flaring events
in the core and these hotspots, it was inferred that an unseen beam
transferred energy between the core and the lobes at a velocity
$>0.95c$.  Limits on the jet size in Cir X-1 were derived by
\citet{Pre83} and \citet{Phi07}, and although scatter-broadened, it
was speculated that a compact jet of size 35\,mas ($175(d/5{\rm
  kpc})$\,AU) might be present in the system.  While Sco X-1 is a
Z-source, Cir X-1 shows characteristics of both atoll and Z-sources
\citep{Oos95}, and Aql X-1 is an atoll source, none of them show clear
evidence for internal shocks propagating down the jets at relativistic
speeds and giving rise to discrete radio knots, as seen in BH
outbursts.  While approximate mass scalings would suggest that
optically-thin NS jets should be less bright than those in BH, they
should be detectable, albeit on smaller scales.  Thus the mechanism
responsible for generating internal shocks (variations in the jet
Lorentz factor are often invoked) is either less effective in NS XRBs
than in their BH counterparts, or the shocks are less efficient
emitters.  Possible explanations include a less variable or more
smoothly varying outflow speed \citep[although NS jets can produce
  ultrarelativistic outflows;][]{Fen04b}, a high sound speed (although
with a maximum of $c/\sqrt{3}$, this appears unlikely if all NS jets
are highly relativistic), or a high magnetic field to smooth out
variations in the flow (possible, given the existing stellar field).
More speculatively, should the discrete knots be ballistic ejecta
launched from magnetic fields threading the ergosphere
\citep{Bla77,Pun90}, rather than internal shocks, this mechanism would
not be present in NS systems.  Disk fields \citep{Bla82} would then be
the best candidate for accelerating the NS and hard state BH jets, as
suggested by \citet{Mei03}.

Further sensitive, high-resolution radio observations of NS sources in
outburst are required to confirm this lack of transient ejecta and
ascertain whether this is indeed a fundamental difference between NS
and BH jets.  Similarities and differences between the two classes of
source will constrain the role of the stellar surface and magnetic
field and the depth of the potential well in accelerating relativistic
jets from accreting compact objects.

\section{Conclusions}
We have obtained unprecedented radio coverage of an outburst of the NS
XRB Aql X-1, demonstrating that the radio emission is consistent with
being activated by both transitions from a hard state to a soft state
and by the reverse transition at lower X-ray luminosity.  Our data
appear to confirm previous suggestions of quenched radio emission
above a certain X-ray luminosity, corresponding to $\sim10$\% of the
Eddington luminosity. We have for the first time detected the source
with VLBI, which together with the measured brightness temperature,
radio spectra and integrated flux density observed by the VLA
demonstrates that the radio emission in this source is consistent with
the compact, partially self-absorbed steady jets seen in the hard
states of BH XRBs.  NSs thus appear equally capable of launching jets
as BHs, but the absence of bright, optically thin, relativistically
moving knots may suggest a fundamental difference in the jet formation
process between NS and BH systems, possibly hinting at an additional
role from the ergosphere.

\acknowledgments J.C.A.M.-J.\ is a Jansky Fellow of the National Radio
Astronomy Observatory (NRAO).  M.L. is supported by the Netherlands
Organisation for Scientific Research (NWO).  D.M.R. acknowledges
support from a NWO Veni Fellowship.  The VLA and VLBA are facilities
of the NRAO which is operated by Associated Universities, Inc., under
cooperative agreement with the National Science Foundation.  The
European VLBI Network is a joint facility of the European, Chinese,
South African and other radio astronomy institutes funded by their
national research councils.  e-VLBI developments in Europe were
supported by the EC DG-INFSO funded Communication Network Developments
project `EXPReS'.  Quick-look RXTE results provided by the ASM/RXTE
team.  Swift/BAT transient monitor results provided by the Swift/BAT
team.

{\it Facilities:} \facility{Swift}, \facility{RXTE}, \facility{VLA}, \facility{VLBA}, \facility{EVN}

\footnotetext[1]{\texttt http://swift.gsfc.nasa.gov/docs/swift/results/transients/AqlX-1.lc.txt}
\footnotetext[2]{\texttt http://xte.mit.edu/XTE/asmlc/srcs/aqlx1.html}

\clearpage

\clearpage

\clearpage

\clearpage

\begin{deluxetable}{lccccccccccl}
\tabletypesize{\scriptsize} \rotate \tablecaption{X-ray and radio
observations of Aql X-1.
\label{tab:obs}}
\tablewidth{0pt}
\tablehead{
\colhead{Date} & \colhead{MJD} & \colhead{PCA} & \colhead{Hard} & \colhead{Soft} &
\colhead{MJD} & \colhead{8.4\,GHz} & \colhead{4.8\,GHz} & \colhead{MJD} & \colhead{8.4\,GHz} &
\colhead{5.0\,GHz} & {X-ray}\\
& PCA & flux & color\tablenotemark{a} & color\tablenotemark{b} & VLA & VLA flux & VLA flux & VLBA & VLBA
flux & EVN flux & state\tablenotemark{c}\\
& (day) & (mCrab) & & & (day) & (mJy) & (mJy) & (day) & (mJy) & (mJy)
&}
\startdata
2006 May 28 & \nodata & \nodata & \nodata & \nodata & 53883.57 & \nodata & $<0.17$ & \nodata & \nodata & \nodata & \nodata \\
2006 May 30 & \nodata & \nodata & \nodata & \nodata & 53885.36 & \nodata & $0.46\pm0.08$ & \nodata & \nodata & \nodata & \nodata \\
2006 Aug 02 & 53949.41 & $24.65\pm0.06$ & $0.729\pm0.008$ & $0.958\pm0.006$ & 53949.41 & $0.23\pm0.05$ & $0.25\pm0.06$ & \nodata & \nodata & \nodata & IS\\
2006 Aug 03 & 53950.34 & $33.10\pm0.10$ & $0.621\pm0.009$ & $0.902\pm0.006$ & 53950.21 & $0.30\pm0.04$ & \nodata & \nodata & \nodata & \nodata & IS\\
2006 Aug 04 & 53951.45 & $37.25\pm0.04$ & $0.694\pm0.004$ & $0.930\pm0.003$ & 53951.06 & $0.18\pm0.04$ & \nodata & \nodata & \nodata & \nodata & IS\\
2006 Aug 07 & 53954.20 & $36.91\pm0.05$ & $0.801\pm0.004$ & $0.993\pm0.003$ & 53954.38 & $0.31\pm0.04$ & $0.31\pm0.07$
& \nodata & \nodata & \nodata & IS\\
2006 Aug 09 & 53955.97 & $26.25\pm0.06$ & $0.966\pm0.008$ & $1.074\pm0.006$ & 53956.27 & $<0.11$ & \nodata & \nodata & \nodata & \nodata & IS\\
2006 Aug 11 & 53958.06 & $15.46\pm0.03$ & $1.062\pm0.007$ & $1.024\pm0.005$ & 53958.34 & $<0.08$ & $<0.11$ & \nodata & \nodata & \nodata & EIS\\
2006 Aug 18 & 53965.46 & $7.82\pm0.02$ & $1.119\pm0.013$ & $0.966\pm0.007$ & 53965.28 & $<0.12$ & \nodata & \nodata & \nodata & \nodata & EIS\\
2006 Aug 21 & \nodata & \nodata & \nodata & \nodata & 53968.34 & $<0.14$ & $<0.18$ & \nodata & \nodata & \nodata & \nodata \\
2006 Aug 23 & \nodata & \nodata & \nodata & \nodata & 53970.33 & $<0.14$ & $<0.23$ & \nodata & \nodata & \nodata & \nodata \\
2006 Aug 28 & \nodata & \nodata & \nodata & \nodata & 53975.32 & $<0.12$ & \nodata & \nodata & \nodata & \nodata & \nodata \\
2006 Aug 31 & \nodata & \nodata & \nodata & \nodata & 53978.21 & $<0.10$ & \nodata & \nodata & \nodata & \nodata & \nodata \\
2007 May 22 & 54242.41 & $39.53\pm0.06$ & $1.064\pm0.006$ & $1.156\pm0.005$ & 54242.59 & $<0.24$ & $<0.66$ & \nodata & \nodata & \nodata & EIS\\
2007 Sep 28 & \nodata & \nodata & \nodata & \nodata & 54371.05 & $0.28\pm0.04$ & $0.29\pm0.05$ & \nodata & \nodata & \nodata & \nodata \\
2007 Sep 30 & \nodata & \nodata & \nodata & \nodata & 54373.87 & $0.37\pm0.09$ & \nodata & \nodata \\
2007 Oct 02 & 54376.99 & $13.88\pm0.05$ & $1.081\pm0.014$ & $1.026\pm0.009$ & 54375.99 & $0.16\pm0.05$ & $<0.18$ & \nodata & \nodata & \nodata & EIS\\
2007 Oct 06 & 54379.87 & $9.60\pm0.04$ & $1.116\pm0.019$ & $0.971\pm0.010$ & 54379.17 & $<0.10$ & $<0.12$ & \nodata & \nodata & \nodata & EIS\\
2007 Oct 07 & \nodata & \nodata & \nodata & \nodata & 54380.85 & $<0.18$ & $<0.24$ & \nodata & \nodata & \nodata & \nodata \\
2007 Oct 19 & \nodata & \nodata & \nodata & \nodata & 54392.99 & $<0.14$ & $<0.18$ & \nodata & \nodata & \nodata & \nodata\\
\hline
2009 Nov 05 & 55140.67 & $10.71\pm0.04$ & $1.145\pm0.015$ & $1.002\pm0.011$ & 55140.12 &
$<0.20$ & \nodata & \nodata & \nodata & \nodata & EIS\\
2009 Nov 06 & 55141.38 & $9.48\pm0.03$ & $1.145\pm0.013$ & $1.011\pm0.009$ & 55141.19 &
$<0.21$ & \nodata & \nodata & \nodata & \nodata & EIS\\
2009 Nov 07 & 55142.09 & $7.83\pm0.04$ & $1.154\pm0.019$ & $0.971\pm0.011$
& \nodata & \nodata & \nodata & \nodata & \nodata & \nodata & EIS\\
2009 Nov 08 & 55143.07 & $10.54\pm0.03$ & $1.155\pm0.012$ & $1.019\pm0.008$
& 55142.99 & $<0.12$ & \nodata & \nodata & \nodata & \nodata & EIS\\
2009 Nov 09 & 55144.12 & $28.06\pm0.05$ & $1.121\pm0.007$ & $1.116\pm0.006$
& \nodata & \nodata & \nodata & \nodata & \nodata & \nodata & EIS\\
2009 Nov 09 & 55144.97 & $39.46\pm0.06$ & $1.103\pm0.005$ & $1.162\pm0.005$
& 55144.97 & $<0.13$ & \nodata & \nodata & \nodata & \nodata & EIS\\
2009 Nov 11 & 55146.08 & $43.19\pm0.05$ & $1.103\pm0.004$ & $1.159\pm0.004$
& 55146.09 & $<0.14$ & \nodata & \nodata & \nodata & \nodata & EIS\\
2009 Nov 12 & 55147.14 & $46.97\pm0.13$ & $1.093\pm0.010$ & $1.185\pm0.009$
& \nodata & \nodata & \nodata & \nodata & \nodata & \nodata & EIS\\
2009 Nov 13 & 55148.25 & $54.30\pm0.09$ & $1.082\pm0.006$ & $1.196\pm0.006$
& \nodata & \nodata & \nodata & \nodata & \nodata & \nodata & EIS\\
2009 Nov 14 & 55149.09 & $64.88\pm0.09$ & $1.048\pm0.005$ & $1.171\pm0.005$
& 55149.08 & $<0.30$ & \nodata & \nodata & \nodata & \nodata & EIS\\
2009 Nov 15 & 55150.00 & $67.76\pm0.13$ & $1.040\pm0.006$ & $1.195\pm0.007$
& \nodata & \nodata & \nodata & \nodata & \nodata & \nodata & EIS\\
2009 Nov 15 & 55150.28 & $67.87\pm0.07$ & $1.054\pm0.004$ & $1.192\pm0.003$ & \nodata & \nodata & \nodata & \nodata & \nodata & \nodata & EIS\\
2009 Nov 16 & 55151.19 & $77.27\pm0.06$ & $0.958\pm0.003$ & $1.162\pm0.002$ & \nodata & \nodata & \nodata & \nodata & \nodata & \nodata & IS\\
2009 Nov 17 & 55152.17 & $85.42\pm0.08$ & $0.860\pm0.003$ & $1.095\pm0.003$ & 55152.07 & $0.68\pm0.09$ & \nodata & \nodata & \nodata & \nodata & IS\\
2009 Nov 18 & 55153.08 & $96.99\pm0.11$ & $0.740\pm0.003$ & $0.998\pm0.003$ & \nodata & \nodata & \nodata & \nodata & \nodata & \nodata & IS\\
2009 Nov 19 & 55154.60 & $148.00\pm0.25$ & $0.515\pm0.004$ & $0.900\pm0.004$ & 55154.00 &
$0.42\pm0.05$ & $0.41\pm0.06$ & 55154.94 & $0.39\pm0.05$ & $0.40\pm0.05$\tablenotemark{d} & IS\\
2009 Nov 20 & 55155.04 & $152.59\pm0.10$ & $0.528\pm0.002$ & $0.886\pm0.001$ & 55155.81 &
$0.20\pm0.06$ & $<0.17$ & \nodata & \nodata & \nodata & BS\\
2009 Nov 21 & 55156.09 & $220.63\pm0.13$ & $0.490\pm0.001$ & $0.992\pm0.001$ & 55156.82 & $<0.08$ & \nodata & \nodata & \nodata & \nodata & BS\\
2009 Nov 22 & 55157.14 & $227.52\pm0.11$ & $0.507\pm0.001$ & $0.975\pm0.001$ & \nodata & \nodata & \nodata & 55157.12 &
$<0.32$ & \nodata & BS\\
2009 Nov 22 & 55157.85 & $189.97\pm0.20$ & $0.503\pm0.002$ & $0.923\pm0.002$ & \nodata & \nodata & \nodata & 55157.86 &
$0.23\pm0.06$ & \nodata & BS\\
2009 Nov 23 & 55158.97 & $280.73\pm0.21$ & $0.498\pm0.002$ &
$1.055\pm0.002$ & \nodata & \nodata & \nodata & \nodata & \nodata & \nodata & BS\\
2009 Nov 24 & \nodata & \nodata & \nodata & \nodata & 55159.10 & $<0.13$ & \nodata & \nodata & \nodata & \nodata & BS\\
2009 Nov 25 & 55160.15 & $244.67\pm0.20$ & $0.497\pm0.002$ & $0.985\pm0.002$ & \nodata & \nodata & \nodata & \nodata & \nodata & \nodata & BS\\
2009 Nov 26 & 55161.07 & $251.19\pm0.14$ & $0.496\pm0.001$ & $0.945\pm0.001$ & \nodata & \nodata & \nodata & 55161.06 & $<0.19$ & \nodata & BS\\      
2009 Nov 27 & 55162.05 & $260.50\pm0.14$ & $0.504\pm0.001$ & $0.956\pm0.001$ & \nodata & \nodata & \nodata & 55162.94 & $<0.14$ & \nodata & BS\\
2009 Nov 28 & 55163.03 & $254.56\pm0.14$ & $0.498\pm0.001$ & $0.975\pm0.001$ & \nodata & \nodata & \nodata & \nodata & \nodata & \nodata & BS\\
2009 Nov 28 & 55163.94 & $245.53\pm0.13$ & $0.486\pm0.001$ & $0.984\pm0.001$ & 55163.94 & $<0.10$ & \nodata & \nodata & \nodata & \nodata & BS\\
2009 Nov 30 & 55165.12 & $164.62\pm0.11$ & $0.534\pm0.002$ & $0.891\pm0.001$ & \nodata & \nodata & \nodata & \nodata & \nodata & \nodata & BS\\      
2009 Dec 01 & 55166.10 & $156.87\pm0.10$ & $0.527\pm0.002$ & $0.917\pm0.001$ & \nodata & \nodata & \nodata & \nodata & \nodata & \nodata & BS\\
2009 Dec 01 & 55166.88 & $138.99\pm0.11$ & $0.500\pm0.002$ & $0.920\pm0.002$ & \nodata & \nodata & \nodata & \nodata & \nodata & \nodata & BS\\
2009 Dec 03 & 55168.06 & $101.59\pm0.09$ & $0.497\pm0.002$ & $0.896\pm0.002$ & \nodata & \nodata & \nodata & \nodata & \nodata & \nodata & BS\\
2009 Dec 05 & 55170.22 & $47.42\pm0.09$ & $0.577\pm0.005$ & $0.934\pm0.004$
& \nodata & \nodata & \nodata & \nodata & \nodata & \nodata & IS\\
2009 Dec 06 & 55171.07 & $32.09\pm0.04$ & $0.863\pm0.004$ & $1.028\pm0.003$ & \nodata & \nodata & \nodata & \nodata & \nodata & \nodata & IS\\
2009 Dec 07 & 55172.05 & $25.72\pm0.04$ & $0.944\pm0.006$ & $1.034\pm0.004$ & 55172.01 & $0.25\pm0.04$ & \nodata & \nodata & \nodata & \nodata & IS\\      
2009 Dec 08 & 55173.03 & $16.86\pm0.03$ & $1.040\pm0.008$ & $1.052\pm0.006$
& 55173.09 & $0.14\pm0.05$ & \nodata & 55173.92 & $<0.22$ & \nodata & EIS\\
2009 Dec 09 & 55174.28 & $11.47\pm0.04$ & $1.115\pm0.013$ & $1.010\pm0.008$ & \nodata & \nodata & \nodata & \nodata & \nodata & \nodata & EIS\\
2009 Dec 10 & 55175.06 & $9.96\pm0.03$ & $1.134\pm0.014$ & $0.990\pm0.008$ & \nodata & \nodata & \nodata & \nodata & \nodata & \nodata & EIS\\
2009 Dec 13 & \nodata & \nodata & \nodata & \nodata & \nodata & \nodata & \nodata & 55178.92 & $<0.17$ & \nodata & EIS
\enddata
\tablenotetext{a}{Count rate ratio (9.7--16.0~keV~/~6.0--9.7~keV)}
\tablenotetext{b}{Count rate ratio (3.5--6.0~keV~/~2.0--3.5~keV)}
\tablenotetext{c}{EIS, IS and BS denote extreme island state, island
state, and banana state, respectively, defined from RXTE observations within 2\,d of the radio data.}
\tablenotetext{d}{The MJD of the EVN observation was 55154.70.}
\end{deluxetable}

\end{document}